 \documentclass[11pt]{article}

  \usepackage{latexsym}
  \usepackage{amssymb}

  \def\AFOUR{%
  \setlength{\textheight}{9.0in}%
  \setlength{\textwidth}{5.75in}

  \setlength{\topmargin}{-0.375in}%
  \hoffset=-.5in%
  \renewcommand{\baselinestretch}{1.17}%
  \setlength{\parskip}{6pt plus 2pt}%
  }

  \AFOUR                                          

  \makeatletter
  \def\section{\@startsection {section}{1}{\z@}{-3.5ex plus -1ex minus
  -.2ex}{2.3ex plus .2ex}{\large\bf}}
  \def\subsection{\@startsection{subsection}{2}{\z@}{-3.25ex plus -1ex minus
  -.2ex}{1.5ex plus .2ex}{\normalsize\bf}}
  \makeatother

  \makeatletter
  \@addtoreset{equation}{section}
  
  \makeatother

\begin{document}
\begin{flushright}
{\it \bf ICMPA-MPA/06/2013}
\end{flushright}
\begin{centering}


{\Large\bf  Separation of spin and charge in the continuum  Schr{o}dinger equation}\\
\vspace{0.5cm}
Dine Ousmane Samary
 \footnote{E-mail: {\tt dine.ousmanesamary@cipma.uac.bj;\,\, ousmanesamarydine@yahoo.fr}}\\
\vspace{0.6cm}

 {\em International Chair of Mathematical
Physics and Applications }\\
{\em Universit\'e d'Abomey Calavi}\\
{\em 072 B.P.: 50 Cotonou, Republic of Benin}\\
\vspace{0.1cm} 
\begin{abstract}
I  describe here the attempt to introduce spin-charge
separation in  Schrodinger equation. The construction we
present here gives a decomposed Schrodinger spinor that has one
problem: Its absolute value can only have value between $0$ and
$\frac{1}{2}$. The problem we solve is to expand and generalize
this construction so that one can have a Schrodinger spinor with
absolute value that are arbitrary non-negative numbers.

It may be that one has to introduce a set of different
decompositions to cover all nonnegative values, that is to
introduce patches over $\mathbb{R}_{+}^{3}$ so that in each patch
one has a different representation.

It seems that the decomposition has a direct relation to so called
entangled states that have been discussed very much in connection
of e.g. quantum computing, and we would like to find this relation
and discuss it in detail.

\end{abstract}


\vspace{0.3pt}

\end{centering}


\setcounter{footnote}{0}

\section{Introduction}
Separation of spin and charge is an behavior of electrons in some materials in which they split into three independent particles ( spinon, orbiton and chargon). It is one of most manifestations of quasiparticles, although that spinon and chargon are note 
gauge invariant quantity.  A spin-charge separation
could have far-reaching practical consequences
to spintronics \cite{spint} that develops devices
which are driven by the spin properties of electrons.
In a wider context~\cite{fad1}, the spin-charge
separation could possibly explain the behavior of elementary
particles in dense environments such as Early Universe
and the interior of compact stars. It might even become
visible in the LHC-ALICE experiment at CERN.
In this  paper  I extend the result on \cite{9} and provide new class of spin and charge decomposition.
\subsection*{Plan of paper}
The work is organized as follows. In section $2$ we discuss the
separation of spin and charge in spin chains and lattice electrons
in the case of non-relativistic context of lattice condensed matter
electrons. In section $3$ we  show  how the previous
lattice  decomposition is realized in the context of the
nonrelativistic Schrodinger quantum mechanics in combination with
Maxwell's electrodynamics. We come to the problem with this
decomposition and show that the decomposition is not complete
because $0\leq |\psi|\leq\frac{1}{2}$. For the Schrodinger
equation we demand that the integral of $|\psi|^2=1$ but at a
given point $x$ the absolute value of the wavefunction can be any
real number. In section $4$ we propose the generalization of the
decomposition and solve the problem of the limit of absolute value
of $\psi$. This generalization shows that $a\leq |\psi|<\infty$,
where $a$ is a real positive value.
 \label{Sect1}
\section{Spin chains and lattice electrons}
In this section we describe spin-charge separation in the
non-relativistic context of lattice condensed matter electrons.

More extensive discussions on spin-charge separation and strongly
correlated lattice electrons can be found e.g. in \cite{1},
\cite{2}, \cite{3}
\subsection{Lattice fermions}
We now describe the quantitative aspects of the spin-charge
separation in a fermionic spin one-half system such as (condensed
matter) electrons.

We consider non-relativistic fermions (electrons) that we describe
by anti-commutating creation and annihilation operators
$c_{l\alpha}^\dag$ and $c_{l\alpha}$, respectively. Here $\alpha$
with the two value $\alpha=\uparrow, \downarrow$ is an index for
the spin-up and spin-down states at the site $l$ of a lattice: We
do not specify the dimensionality of the lattice, in particular it
can now be more than one dimensional. The operators subject to the
anti commutation relations
\begin{eqnarray}\label{1.1}
\{c_{l\alpha},c_{l'\beta}\}=\{c_{l\alpha}^\dag,c_{l'\beta}^\dag\}=0,\,\,\,\,\,
\{c_{l\alpha},c_{l'\beta}^\dag\}=\delta_{ll'}\delta_{\alpha\beta}.
\end{eqnarray}
At each site of the lattice these operators span a Hilbert space
with four states. These are the (Fock) vacuum $|0>$, the state
with spin-up ($\uparrow$) and the state with spin-down
($\downarrow$),
\begin{eqnarray}
c_{\uparrow}^\dag|0>=|\uparrow>,\,\,\,\,
c_{\downarrow}^\dag|0>=|\downarrow>
\end{eqnarray}
and the doubly occupied state
\begin{eqnarray}
c_{\uparrow}^\dag c_{\downarrow}^\dag|0>=|\uparrow\downarrow>.
\end{eqnarray}
The Fock vacuum and the doubly occupied state are bosonic, the
other two are fermionic.

We introduce a spin-charge decomposition in these creation and
annihilation operators. It is defined by the following
transformation
\begin{eqnarray}\label{1.2}
c_{l\alpha}^\dag=b_l. s_{l\alpha}^\dag+d_l^\dag
.\epsilon_{\alpha\beta}s_{l\beta},\,\,\, c_{l\alpha}=b_l^\dag.
s_{l\alpha}+d_l .\epsilon_{\alpha\beta}s_{l\beta}^\dag
\end{eqnarray}
with $\epsilon_{\alpha\beta}=-\epsilon_{\beta\alpha}$ and
$\epsilon_{12}=1$. Here $s_{l\alpha}^\dag$ and $s_{l\alpha}$ are
fermionic creation and annihilation operators. They are called
spinon operators. The spinons clearly carry the spin degree of
freedom in the decomposition. The two bosonic operators $b_l^\dag$
and $b_l$ correspond to states that are called holons. The two
bosonic operators $d_l^\dag$ and $d_l$ correspond to states that
are called doublons.

By construction, the bosonic operators have no spin. But they do
carry the electric charge which becomes apparent when we consider
the symmetry properties of the decomposition. For this we
introduce a local Maxwell $U(1)$ gauge transformation. It acts on
the electron operators in the standard fashion as follows,
\begin{eqnarray}
c_{l\alpha}^\dag\rightarrow e^{i\phi_l}.c_{l\alpha}^\dag
\end{eqnarray}
and
\begin{eqnarray}
c_{l\alpha}\rightarrow e^{-i\phi_l}.c_{l\alpha}.
\end{eqnarray}
From (\ref{1.2}) we conclude immediately that we can take
\begin{eqnarray}\label{1.3}
b_l\rightarrow e^{i\phi_l}b_l,\,\,\, d_l\rightarrow
e^{-i\phi_l}d_l,\,\,\, s_{l\alpha}\rightarrow s_{l\alpha}.
\end{eqnarray}
This implies that the holons and the doublons carry an electric
charge which is equal to that of an electron, while the spinons
are charge neutral. As a consequence (\ref{1.2}) indeed decomposes
the electron creation and annihilation operators in terms of
independent spin and charge carriers.

If we substitute the decomposition (\ref{1.2}) in the
anticommutation relations (\ref{1.1}) of $c_{l\alpha}$ and
$c_{l\alpha}^\dag$ and assume that the holon, doublon and spinon
operators each obey standard canonical anti commutatation
relations of bosons and fermions respectively, we find that the
decomposed electron operator is also subject to the canonical anti
commutators but with the following modification
\begin{eqnarray}\label{1.4}
\{c_{l\alpha},c_{l'\beta}^\dag\}=\Big(b_l^\dag.b_{l'}+d_{l'}^\dag
.d_l+s_{l\uparrow}^\dag .s_{l'\uparrow}+s_{l\downarrow}^\dag
.s_{l'\downarrow}\Big)\delta_{ll'}\delta_{\alpha\beta}.
\end{eqnarray}
Here
\begin{eqnarray}\label{1.5}
\hat{N}_l=b_l^\dag.b_{l'}+d_{l'}^\dag .d_l+s_{l\uparrow}^\dag
.s_{l'\uparrow}+s_{l\downarrow}^\dag .s_{l'\downarrow}
\end{eqnarray}
is the total number operator at the site $l$. In the subspace of
states $|phys>$ that obey
\begin{eqnarray}\label{1.6}
\hat{N}_l|phys>=|phys>
\end{eqnarray}
for every $l$, the decomposed electron creation and annihilation
operators then reproduce the canonical anti commutators
(\ref{1.1}).

The constraint (\ref{1.6}) states that exactly one particle
occupies each lattice site. In this sense the liquid composed of
spinons and chargons is incompressible.

We also note that the constraint (\ref{1.6}) resembles the Gaub
law constraint of a gauge theory, with $\hat{N}_l$ the (abelian)
charge operator.

One can show that in the subspace of states $|phys>$ the holon
operators can be interpreted as creation and annihilation
operators corresponding to the bosonic and spinless Fock vacuum
state $|0>$ of the electron, the doublons as creation and
annihilation operators for the spinless and bosonic doubly
occupied state $|\uparrow\downarrow>$, and the spinons as creation
and annihilation operators corresponding to the fermionic spin
$\pm 1/2$ states $|\uparrow>$ and $|\downarrow>$.

Beside the action of the Maxwellian $U(1)$ phase rotation, we also
identify in the decomposition (\ref{1.2}) the following local
internal $U_{int}(1)$ symmetry: if we send
\begin{eqnarray}\label{1.7}
b_l\rightarrow e^{i\eta_l}b_l,\,\,\, d_l\rightarrow
e^{i\eta_l}d_l,\,\,\, s_{l\alpha}\rightarrow
e^{i\eta_l}s_{l\alpha}
\end{eqnarray}
the electron operators $c_{l\alpha}$ and $c_{l\alpha}^\dag$ remain
intact. Since the decomposition should not introduce any
(vortex-like) line singularities in the operators, we define the
angle $\eta_l$ modulo $2\pi$. This implies that the internal
$U_{int}(1)$ gauge group is compact. This internal $U(1)$ symmetry
is presumed to have important physical consequences. In
particular, since $U_{int}(1)$ is a compact group we expect
\cite{4} that there is a first-order phase transition between a
strong coupling confined phase and a week coupling deconfined
phase. Since there is a priori no natural kinetic term for any
dynamically independent gauge field of the internal group, we are
formally in the infinite coupling limit with respect to a coupling
of the internal group. This means that under normal circumstances
we are in the confinement phase of the compact gauge theory. From
the point of view of physical applications this could be the
explanation why under these normal circumstances the spinon and
the chargon are tightly confined to each other into a pointlike
electron.

In the minimal version of the compact $U(1)$ theory the coupling
does not run. But it is  conceivable that in the presence of
additional fields such as the spin and charge variables, the
$\beta-$function for the coupling becomes nontrivial. Since the
gauge group is Abelian, we expect  that the strength of the
coupling increases at high energies. This leads us to a picture
which is consistent with the known pointlike structure of an
individual $S-$matrix electron in the high energy limit. The
shorter the distance, the tighter the chargon and holon become
confined to each other. But at low energies and in a proper
quantum protectorate where the coupling of the internal gauge
group becomes weak the spinon and chargons may become deconfined.
A spin-charge decomposition can then take place and instead of
pointlike electrons, fundamental constituents are deconfined
spinons and chargons.

Suppose that the internal gauge symmetry corresponds to a
fundamental interaction akin Maxwell's electromagnetism, that
there is indeed a yet unseen fundamental force which under normal
material conditions is so strong that it confines the spinons and
chargons into pointlike electrons. The condensed matter electron
would like the Cooper pair of spinon and chargon, and the
different quantum protectorates would describe different phases in
the conventional sense, separated from each other by conventional
phase transitions of the internal gauge group. The concept of a
quantum protectorate would then reduce to the standard concept of
a phase. But until now, there is no experimental evidence of a new
fundamental force corresponding to the compact internal $U(1)$.
Thus this internal symmetry remains a mathematical construct and
the notion of a quantum protectorate endures as conceptually
distinct from the notion of a phase.

The decomposition (\ref{1.2}) is not unique. For example, we could
also introduce the truncated decomposition
\begin{eqnarray}\label{1.8}
c_{l\alpha}^\dag=b_l s_{l\alpha}^\dag,\,\,\,\,
c_{l\alpha}=b_l^\dag s_{l\alpha}
\end{eqnarray}
Now the canonical fermionic anti commutation relations lead to the
following structure (we suppress the lattice site index $l$)
\begin{eqnarray}\label{1.9}
\{c_\uparrow^\dag, c_\uparrow\}=b^\dag b+s_\uparrow^\dag
s_\uparrow=:\hat{N}_\uparrow
\end{eqnarray}
\begin{eqnarray}\label{1.10}
\{c_\downarrow^\dag, c_\downarrow\}=b^\dag b+s_\downarrow^\dag
s_\downarrow=:\hat{N}_\downarrow
\end{eqnarray}
\begin{eqnarray}\label{1.11}
\{c_\uparrow^\dag,c_\downarrow\}=s_\uparrow^\dag
s_\downarrow=:\hat{\mathcal{C}^-}
\end{eqnarray}
\begin{eqnarray}\label{1.12}
\{c_\downarrow^\dag,c_\uparrow\}=s_\downarrow^\dag
s_\uparrow=:\hat{\mathcal{C}^+}
\end{eqnarray}
 with all other anticommutators vanishing. The relation (\ref{1.9}), (\ref{1.10}) (\ref{1.11}) (\ref{1.12}) can be generalized as
\begin{eqnarray}
\{c_\alpha^\dag, c_\beta\}=s_\alpha^\dag s_\beta+ b^\dag b
\delta_{\alpha\beta}
\end{eqnarray}
 As a consequence the
decomposed operators (\ref{1.8}) reproduce the canonical fermionic
anticommutation relations in the subspace of states $|phys>$ that
are subject to the conditions
\begin{eqnarray}\label{1.13}
\hat{N}_\downarrow |phys>=\hat{N}_\downarrow |phys>=|phys>
\end{eqnarray}
and
\begin{eqnarray}\label{1.14}
\hat{\mathcal{C}^\pm}|phys>=0
\end{eqnarray}
and in addition the (anti)commutators of these four operators must
vanish weakly, in the subspace of states $|phys>$.

We first observe that
\begin{eqnarray}
[\hat{\mathcal{C}^-},\hat{\mathcal{C}^+}]=s_\uparrow^\dag
s_\uparrow-s_\downarrow^\dag s_\downarrow=:\hat{\mathcal{C}^0}
\end{eqnarray}
and
\begin{eqnarray}
[\hat{\mathcal{C}^+},\hat{\mathcal{C}^0}] &=&s_\downarrow^\dag
s_\uparrow \Big(2s_\uparrow^\dag s_\uparrow-2s_\downarrow^\dag
s_\downarrow\Big)=2\hat{\mathcal{C}^+}\hat{\mathcal{C}^0}
\end{eqnarray}
\begin{eqnarray}
[\hat{\mathcal{C}^-},\hat{\mathcal{C}^0}] &=&s_\uparrow^\dag
s_\downarrow \Big(2s_\uparrow^\dag s_\uparrow
 -2 s_\downarrow^\dag s_\downarrow\Big)=2\hat{\mathcal{C}^-}\hat{\mathcal{C}^0}
\end{eqnarray}
consequently in the constraint surface
$(\hat{\mathcal{C}^0},\hat{\mathcal{C}^\pm})$ defines a $SU(2)$
algebra and consistency of our construction demands that the
states $|phys>$ are subject to the (first class) $SU(2)$
constraint algebra (Gaub law).
\begin{eqnarray}
\hat{\mathcal{C}^\pm}|phys>=\hat{\mathcal{C}^0}|phys>=0.
\end{eqnarray}
Furthermore, since we clearly have
\begin{eqnarray}
[\hat{\mathcal{C}^\pm},\hat{N}_\uparrow]=[\hat{\mathcal{C}^\pm},
\hat{N}_\downarrow]=[\hat{\mathcal{C}^0},\hat{N}_\uparrow]=[\hat{\mathcal{C}^0},\hat{N}_\downarrow]=0
\end{eqnarray}
the constraint algebra is also consistent with the conditions
(\ref{1.13}). This ensures that in the subspace $|phys>$ the
decomposed operators do realize the canonical fermionic anti
commutation relations.

Finally, the states $|phys>$ can be constructed as follows: We
start from the following Fock ground state $|0>$,
\begin{eqnarray}
b|0>=s_\uparrow|0>=s_\downarrow|0>=0.
\end{eqnarray}
Clearly, this state also obeys
\begin{eqnarray}
\hat{\mathcal{C}^\pm}|0>=\hat{\mathcal{C}^0}|0>=0.
\end{eqnarray}
We then introduce the following two states
\begin{eqnarray}
|b>=b^\dag|0>,\,\,\,\, |\uparrow\downarrow>=s_\uparrow^\dag
s_\downarrow^\dag|0>.
\end{eqnarray}
One can easily verify that these two states satisfy the conditions
we have imposed on $|phys>$. These two states span the Hilbert
space where the decomposed operators (\ref{1.8}) realize the
fermionic anti commutation relation.

Note that on the remaining two states at each lattice site,
\begin{eqnarray}
|\uparrow>=s_\uparrow^\dag|0>,\,\,\,\,
|\downarrow>=s_\downarrow^\dag |0>
\end{eqnarray}
the action of the constraint operators
$(\hat{\mathcal{C}^0},\hat{\mathcal{C}^\pm})$ becomes realized in
terms of conventional Pauli matrices.

Clearly, the physical state $|0>$ can be identified as the holon,
and the physical state $|\uparrow\downarrow>$ can be identified as
the doublon. Consequently (\ref{1.8}) is a projection of the
electron creation and annihilation operators to the subspace of
chargons.

Finally, there is also the following decomposition \cite{3}.
\begin{eqnarray}\label{1.17}
c_{l\alpha}^\dag=\frac{b_l^\dag}{\sqrt{b_l^\dag
b_l}}.s_{l\alpha}^\dag,\,\,\,\,\,
c_{l\alpha}=\frac{b_l}{\sqrt{b_l^\dag b_l}}.s_{l\alpha}.
\end{eqnarray}
These decomposed operators satisfy the fermionic anti commutation
relations without any additional constraints.

A decomposition such as (\ref{1.2}), (\ref{1.8}) and (\ref{1.17})
admits a well-defined interpretation in terms of group
representation theory: the original creation and annihilation
operators realize a tensor product representation of the
Maxwellian gauge group and the spatial rotation group. The
spin-charge operators can be interpreted in terms of a
Clebsch-Gordan expansion that decomposes this tensor product into
irreducible representations of the two groups.

Usually in Physics, we have learned to expect that the irreducible
components in a Glebsch-Gordan expansion have more fundamental
physical value than their reducible tensor products. But in the
present case the physical relevance of the decomposition derives
from two added dynamical criteria:
\begin{itemize}
\item The decomposition must be consistent with the particle
interpretation of the operators involved. This means that
(anti)commutation relations of all operators involved must have
the canonical structure which is consistent with the creation and
annihilation of individual particle states.

\item When the Hamiltonian is realized in terms of the decomposed
operators, it must have a physically meaningful form.

Only when these two added criteria are fulfilled, we can expect
that a spin-charge separation becomes realized in the Nature. From
our schematic examples we expect that this will be the case at
least when we implement a spin-charge decomposition in a proper
Hamiltonian realization of the one dimensional Neel
antiferromagnet.
\end{itemize}
\subsection{Hubbard Model}
Nowadays, there is little doubt about the microscopic theory that
describes cuprate superconductors \cite{1}. These materials are
modeled by a (single band) Hubbard-model \cite{5} with Hamiltonian
of the form
\begin{eqnarray}
H=H_t+H_U=-\sum_{l,l',\alpha}t_{l-l'}c_{l\alpha}^\dag
c_{l\alpha}+U\sum_{l}n_{l\uparrow}n_{l\downarrow}.
\end{eqnarray}
Here the summations extend over all ($2d$) lattice sites $l$ and
over the spin degrees of freedom $\alpha= \uparrow,\downarrow$ and
\begin{eqnarray}
n_{l\alpha}=c_{l\alpha}^\dag c_{l\alpha}
\end{eqnarray}
is the number operator and $t_{l-l'}$ and $U$ are phenomenological
parameters.

The properties of the Hubbard model including spin-charge
separation have been discussed widely in the literature, and we
refer to \cite{1},  \cite{2},  \cite{5} for details.

When we introduce the spin operator
\begin{eqnarray}\label{1.18}
{\bf \mathcal{S}}_l=\frac{1}{2}\sum_{\alpha\beta}c_{l\alpha}^\dag
\hat{\sigma}_{\alpha\beta}c_{l\beta}
\end{eqnarray}
where
\begin{eqnarray}
\hat{\sigma}=(\sigma^1,\sigma^2, \sigma^2)
\end{eqnarray}
are the three Pauli matrices, we can write
\begin{eqnarray}
H_U=\frac{U}{2}\sum_{l}(n_{l\uparrow}+n_{l\downarrow})-\frac{2U}{3}\sum_{l}{\bf
\mathcal{S}}_l^2
\end{eqnarray}
and if we demand the strong repulsion constraint
\begin{eqnarray}\label{1.19}
\sum_{\alpha}c_{l\alpha}^\dag c_{l\alpha}=1
\end{eqnarray}
the components of the spin operator (\ref{1.18}) satisfy the spin
commutation relations of the Pauli matrices.

The condition (\ref{1.19}) imposes the constraint that exactly
half of the band is filled. If we remove electrons from the
half-filled state the system becomes underdoped and the constraint
(\ref{1.19}) is replaced by the non-holonomic
\begin{eqnarray}\label{1.20}
\sum_{\alpha}c_{l\alpha}^\dag c_{l\alpha}\leq 1.
\end{eqnarray}
There are now holes in the otherwise uniform charge distribution.
As in our schematic example, these holes which correspond to sites
that are unoccupied by electrons propagate along the lattice when
electrons jump from an occupied site to an unoccupied site. This
leads to the propagation of charge but with no transport of spin.
As a consequence in an underdoped system we expect that the spin
and the charge can be transported independently. Instead of
electrons the fundamental constituents are now quasiparticles that
are either chargeless fermionic spinons or spinless bosonic
holons, which in general travel with different velocities.

A cuprate superconductors is supposedly described by such an
underdoped system. The constraint (\ref{1.20}) is non-holonomic
but it becomes elegantly resolved in terms of the decomposed
variables (\ref{1.2}). For this we simply project the
decomposition (\ref{1.2}) to the subspace where
\begin{eqnarray}\label{1.21}
d_l|phys>=d_l^\dag|phys>=0.
\end{eqnarray}
This clearly removes the doubly occupancy of the states. For
simplicity one may wish to use the truncated decomposition
(\ref{1.8}) in an effective manner, but with proper care since the
constraint structures are different: A naive use of (\ref{1.8})
implements a projection to the subspace of holons and doublons
instead.

In a mean field approximation where we integrate over the fermions
$s_{l\alpha}$ and $s_{l\alpha}^\dag$, we now expect to get
(d-wave) superconductivity when the holon fields $b_l$ condense,
\begin{eqnarray}\label{1.22}
<b_l^\dag b_l>=\Delta_b \neq 0
\end{eqnarray}
which corresponds to the relation (\ref{1.20}) in the present
case.

The ensuing phase diagram is quite elaborate, there are several
different phase regions \cite{1}. Besides (\ref{1.22}), of
particular interest in the pseudo-gap phase with the characteristic
property that even though the underlying symmetry is broken the
effective bosonic order parameter $\Delta_b$ vanishes due to
quantum fluctuations.

Finally, our one dimensional example (figures 3-9) reveals that
there is an apparent symmetry between the transport of holons and
doublons. Obviously, the condition for overdopping can be
introduced in manner which is parallel to (\ref{1.21}), by
projecting to the subspace where
\begin{eqnarray}\label{1.23}
b_l|phys>=b_l^\dag|phys>=0
\end{eqnarray}
and assuming that there is a doublon condensation.

 \label{Sect2}
\section{Spin charge separation and the Schrodinger equation}
In the continuum limit the interactive dynamics of condensed
matter electrons is governed by nonrelativistic Schrodinger
quantum mechanics in combination with Maxwell's electrodynamics.
In this section we study how the previous lattice decompositions
become realized in that context. We find that the spin-charge
decomposed two-component Pauli spinor has a structure which is
very similar to that of the lattice fermion, including the
appearance of a $SU(2)$ constraint algebra.
\subsection{Decomposing the Pauli spinor}
In obvious notation the classical Schrodinger Lagrangian including
the Maxwellian contribution is
\begin{eqnarray}\label{2.1}
\mathcal{L}[\psi,\psi^\dag,A_\mu ]=\psi^\dag\Big(i\partial_0
-eA_0\Big)\psi+\frac{1}{2m}|\Big(i\partial_k-eA_k\Big)|^2-\frac{1}{4}F_{\mu\nu}^2.
\end{eqnarray}
Here $\psi$ is the two-component commuting (Pauli) spinor. We
interpret it as a Hartree-type many-body wavefunction that
describes the nonrelativistic dynamics of an ensemble of
interacting electrons in its totally antisymmetric subspace
\cite{6}.

From (\ref{2.1}) we infer the following familiar canonical Poisson
brackets for the Pauli spinor
\begin{eqnarray}\label{2.2}
\{\psi_\alpha^{\star}(x),\psi_\beta(x')\}=\delta_{\alpha\beta}(x-x');\,\,\,
\{\psi_\alpha^{\star}(x),\psi_\beta^{\star}(x')\}=\{\psi_\alpha(x),\psi_\beta(x')\}=0.
\end{eqnarray}
The structure and normalization of these brackets ensure that at
the level of the quantum theory where the brackets become replaced
by (anti)commutators, the Pauli spinors acquire the standard
interpretation in terms of proper fermion creation and
annihilation operators.

The obvious Ansatz for a spin-charge decomposition is
\begin{eqnarray}\label{2.3}
\psi_\alpha=\phi. \mathcal{S}_\alpha
\end{eqnarray}
where $\phi(x)$ is to be viewed as the chargon and
$\mathcal{S}_\alpha$ as the spinon. However, despite its
naturalness this decomposition leads to difficulties: Even though
(\ref{2.3}) can be viewed as a change of variables in the
Lagrangian once we first remove the degeneracy in the right hand
side, somewhat unexpectedly in turns out that (\ref{2.3}) is not
consistent with the interpretation of the chargon and the spinon
in terms of particle states in the second quantized quantum
theory. This can be seen either by substituting (\ref{2.3}) in
(\ref{2.1}) and carefully analysing the properties of the ensuing
decomposed Lagrangian, or more directly by inspecting the
canonical structure of (\ref{2.3}). Here we follow the latter
route as it is an independent model  and thus it has more generality.

In order that $\phi$ and $\mathcal{S}_\alpha$ acquire a proper
particle interpretation at the quantum level, their classical
Poisson brackets should have the following structure:
\begin{eqnarray}
\{\phi^{\star}(x),
\phi(x')\}=\delta(x-x');\,\,\,\{\mathcal{S}^\star_\alpha(x),\mathcal{S}_\beta(x')\}=\delta_{\alpha\beta}(x-x')
\end{eqnarray}
with all other brackets vanishing. These Poisson brackets ensure
that at the second quantized level we can consistently expand the
fields $\phi(x)$ and $\mathcal{S}_\alpha$ in terms of particle
creation and annihilation operators, in the usual manner.

If we substitute the decomposition (\ref{2.3}) in (\ref{2.2}), we
arrive at the following non-vanishing brackets for the decomposed
Pauli spinor,
\begin{eqnarray}
\{\psi_\alpha^\star(x), \psi_\alpha(x')\}
&=&\Big(\phi^\star(x)\phi(x)+\mathcal{S}_\alpha^\star(x)\mathcal{S}_\alpha(x)\Big)\delta(x-x')\cr
&=&\varrho_\alpha(x) \delta(x-x'),\,\,\, (\alpha=1,2)
\end{eqnarray}
and
\begin{eqnarray}
\{\psi_1^\star(x), \psi_2(x')\}
&=&\mathcal{S}_1^\star(x)\mathcal{S}_2(x)\delta(x-x')=:\mathcal{C}^-(x)\delta(x-x')
\end{eqnarray}
\begin{eqnarray}
\{\psi_2^\star(x), \psi_1(x')\}
&=&\mathcal{S}_2^\star(x)\mathcal{S}_1(x)\delta(x-x')=:\mathcal{C}^+(x)\delta(x-x').
\end{eqnarray}
These brackets reproduce (\ref{2.2}) provided we can consistently
impose the constraints,
\begin{eqnarray}\label{2.4}
\varrho_\alpha(x)\approx 1,\,\,\, (\alpha=1,2),\,\,\,\mbox{ and
}\,\,\,\mathcal{C}^\pm(x)\approx 0
\end{eqnarray}
These constraint functionals have the following non-vanishing
Poisson bracket relations
\begin{eqnarray}
\{\mathcal{C}^-(x),
\varrho_1(x')\}&=&\mathcal{S}_1^\star(x)\mathcal{S}_2(x)\delta(x-x')
=\mathcal{C}^-(x)\delta(x-x')
\end{eqnarray}
\begin{eqnarray}
\{\mathcal{C}^+(x),
\varrho_1(x')\}&=&-\delta(x-x')\mathcal{S}_2^\star(x)\mathcal{S}_1(x)=-\mathcal{C}^+(x)\delta(x-x')
\end{eqnarray}
and
\begin{eqnarray}
\{\mathcal{C}^-(x),
\varrho_2(x')\}&=&-\mathcal{S}_1^\star(x)\mathcal{S}_2(x)\delta(x-x')
=-\mathcal{C}^-(x)\delta(x-x')
\end{eqnarray}
\begin{eqnarray}
\{\mathcal{C}^+(x),
\varrho_2(x')\}&=&\mathcal{S}_2^\star(x)\mathcal{S}_1(x)\delta(x-x')=\mathcal{C}^+(x)\delta(x-x').
\end{eqnarray}
Then
\begin{eqnarray}
\{\mathcal{C}^\pm(x),
\varrho_1(x')\}=\mp\mathcal{C}^\pm(x)\delta(x-x'),\,\,\,\{\mathcal{C}^\pm(x),
\varrho_2(x')\}=\pm\mathcal{C}^\pm(x)\delta(x-x').
\end{eqnarray}
We have also
\begin{eqnarray}
\{\mathcal{C}^-(x),\mathcal{C}^+(x')\}
&=&\Big(\mathcal{S}_2^\star(x)\mathcal{S}_2(x)-\mathcal{S}_1^\star(x)\mathcal{S}_1(x)\Big)\delta(x-x')\cr
&=:&\mathcal{C}^{0}(x)\delta(x-x').
\end{eqnarray}
and
\begin{eqnarray}
\{\mathcal{C}^+(x),\mathcal{C}^0(x')\}&=&2\mathcal{S}_2^\star(x')\mathcal{S}_1(x)\delta(x-x')=2\mathcal{C}^+(x)\delta(x-x')
\end{eqnarray}
\begin{eqnarray}
\{\mathcal{C}^-(x),\mathcal{C}^0(x')\}&=&-2\mathcal{S}_1^\star(x)\mathcal{S}_2(x)\delta(x-x')=-2\mathcal{C}^-(x)\delta(x-x').
\end{eqnarray}
We conclude that this gives us the additional non-vanishing
brackets
\begin{eqnarray}
\{\mathcal{C}^\pm(x),\mathcal{C}^0(x')\}=\pm 2 \mathcal{C}^\pm(x).
\end{eqnarray}
In particular, $(\mathcal{C}^\pm(x),\mathcal{C}^0(x))$ determines
an $SU(2)$ algebra.

Combining these Poisson brackets we conclude that we have a
first-class constraint algebra in the sense of Dirac \cite{8},
provided we impose the additional constraint
\begin{eqnarray}\label{2.5}
\mathcal{C}^0(x)\approx 0.
\end{eqnarray}
We now resolve this constraint algebra. We first note that
(\ref{2.5}) gives
\begin{eqnarray}
\mathcal{S}_1^\star(x)\mathcal{S}_1(x)\approx\mathcal{S}_2^\star(x)\mathcal{S}_2(x)
\end{eqnarray}
on the constraint surface. When we combine this with (\ref{2.4})
we conclude that on the constraint surface
\begin{eqnarray}
\mathcal{S}_1(x)\approx \mathcal{S}_2(x)\approx 0.
\end{eqnarray}
Thus the spinon is weakly zero on the constraint surface, and only
the phase of the chargon survives. In particular, the entire Pauli
spinor vanishes weakly on the constraint surface.

We conclude that even though (\ref{2.3}) is an appealing and
natural choice for the spin-charge decomposition Pauli spinor, it
does not lead to a consistent particle interpretation of the
spinon and chargon when we replace the Poisson brackets with
(anti)commutators.

We now show how the spin-charge separation can be imposed in a
manner which is consistent with the particle interpretation. For
this we recall (\ref{1.2}) and introduce the following decomposed
Pauli spinor \cite{9} (we denote $\alpha,\beta=1,2$)
\begin{eqnarray}\label{2.6}
\psi_\alpha(x)=\phi_+(x)\mathcal{S}_{+\alpha}(x)+\phi_-(x)\mathcal{S}_{-\alpha}(x).
\end{eqnarray}
Here $\phi_{\pm}$ are two complex functions and
$\mathcal{S}_{\pm}$ are two a priori linearly independent complex
spinors.

We first argue that (\ref{2.6}) can be viewed as a decomposition
of the Schrodinger wave function into its independent spin and
charge constituents. For this we implement a Maxwellian $U(1)$
gauge transformation that sends the Pauli spinor into
\begin{eqnarray}\label{2.7}
\psi(x)\rightarrow e^{i\eta}\psi(x).
\end{eqnarray}
This suggests \cite{9} that for the decomposed fields we take
\begin{eqnarray}\label{2.8}
\phi_{\pm}\rightarrow e^{i\eta}\phi_{\pm},\,\,\,\,
\mathcal{S}_{\pm}\rightarrow\mathcal{S}_{\pm}.
\end{eqnarray}
Consequently $\phi_{\pm}$ become  chargons that carry the
electric charge of the Pauli spinor. The $\mathcal{S}_{\pm}$ are
charge neutral. But since $\mathcal{S}_{\pm}$ are spinors, they are
the spinons that carry the spin of the electron \cite{10}.

In addition of (\ref{2.8}), we have the following two internal
compact $U_{int}^\pm(1)$ gauge symmetries in the decomposition:
\begin{eqnarray}\label{2.9}
\phi_{\pm}\rightarrow
e^{i\gamma_{\pm}}\phi_{\pm},\,\,\,\,\mathcal{S}_{\pm}\rightarrow
e^{i\gamma_{\pm}}\mathcal{S}_{\pm},\,\,\,\, \psi\rightarrow\psi
\end{eqnarray}
These internal rotations are akin the internal gauge symmetry
(\ref{1.7}). We propose that in analogy with (\ref{1.7}) the
internal symmetry transformation (\ref{2.9}) ensures that under
normal circumstances the spinon and the chargon are confined into
the Pauli spinor $\psi$.

We now proceed to inspect the Poisson bracket structure of the
decomposed spinor.

For the chargons $\phi_{\pm}$, the consistency with the particle
interpretation at the second quantized level proposes that we
postulate the non-vanishing Poisson brackets
\begin{eqnarray}\label{2.10}
\{\phi_{\pm}^\star(x), \phi_{\pm}(x')\}=\delta(x-x').
\end{eqnarray}
Similarly, we assume that the spinon components are subject to the
non-vanishing brackets
\begin{eqnarray}\label{2.11}
\{\mathcal{S}_{\pm\alpha}^\star,\mathcal{S}_{\pm\beta}\}=\delta_{\alpha\beta}(x-x').
\end{eqnarray}
These Poisson bracket relations ensure that the decomposed spinor
(\ref{2.6}) satisfies
\begin{eqnarray}
\{\psi_\alpha(x),\psi_\beta(x')\}=0,\,\,\,\,\,\{\psi_\alpha^\star(x),\psi_\beta^\star(x')\}=0.
\end{eqnarray}
When we compute the remaining Poisson brackets of the decomposed
spinors we arrive at the following non vanishing brackets,
\begin{eqnarray}\label{2.12}
\{\psi_1^\star(x),\psi_1(x')\}&=&\Big(\phi_+^\star\phi_++\phi_-^\star\phi_-
+\mathcal{S}_{+1}^\star\mathcal{S}_{+1}+\mathcal{S}_{-1}^\star\mathcal{S}_{-1}\Big)(x)\delta(x-x')\cr
&=:&\varrho_1(x)\delta(x-x')
\end{eqnarray}
\begin{eqnarray}\label{2.13}
\{\psi_2^\star(x),\psi_2(x')\}&=&\Big(\phi_+^\star\phi_++\phi_-^\star\phi_-
+\mathcal{S}_{+2}^\star\mathcal{S}_{+2}+\mathcal{S}_{-2}^\star\mathcal{S}_{-2}\Big)(x)\delta(x-x')\cr
&=:&\varrho_2(x)\delta(x-x')
\end{eqnarray}
\begin{eqnarray}
\{\psi_1^\star(x),,\psi_2(x')\}
&=&\Big(\mathcal{S}_{+1}^\star\mathcal{S}_{+2}+\mathcal{S}_{-1}^\star\mathcal{S}_{-2}\Big)\delta(x-x')=:\mathcal{C}^{-}(x)\delta(x-x')
\end{eqnarray}
\begin{eqnarray}
\{\psi_2^\star(x),\psi_1(x')\}
&=&\Big(\mathcal{S}_{+2}^\star\mathcal{S}_{+1}+\mathcal{S}_{-2}^\star\mathcal{S}_{-1}\Big)\delta(x-x')=:\mathcal{C}^+(x)\delta(x-x').
\end{eqnarray}
The Poisson brackets of $\mathcal{C}^+$ and $\mathcal{C}^-$ give
\begin{eqnarray}
\{\mathcal{C}^+(x),\mathcal{C}^-(x')\}
&=&\Big(\varrho_1(x)-\varrho_2(x)\Big)\delta(x-x')
\end{eqnarray}
and since
\begin{eqnarray}
\{\mathcal{C}^-(x),\mathcal{C}^+(x')\}=\Big(\varrho_2(x)-\varrho_1(x)\Big)\delta(x-x')=:\mathcal{C}^0(x)\delta(x-x')
\end{eqnarray}
and
\begin{eqnarray}
\{\mathcal{C}^+(x),\mathcal{C}^0(x')\}
&=&2\mathcal{C}^+(x)\delta(x-x')
\end{eqnarray}
\begin{eqnarray}
\{\mathcal{C}^-(x),\mathcal{C}^0(x')\}
&=&-2\mathcal{C}^{-}(x)\delta(x-x')
\end{eqnarray}
\begin{eqnarray}
\{\mathcal{C}^\pm(x),\mathcal{C}^0(x')\}=\pm
2\mathcal{C}^\pm(x)\delta(x-x')
\end{eqnarray}
the  constraint functionals
($\mathcal{C}^0(x),\mathcal{C}^\pm(x)$) determine a $SU(2)$
algebra. Finally, since
\begin{eqnarray}
\{\mathcal{C}^+(x),\varrho_1(x')\} &=&\mathcal{C}^+(x)\delta(x-x')
\end{eqnarray}
\begin{eqnarray}
\{\mathcal{C}^-(x),\varrho_1(x')\}
&=&\mathcal{C}^-(x)\delta(x-x'),
\end{eqnarray}
we conclude that if we define the total density operator
\begin{eqnarray}\label{2.16}
\varrho=\frac{1}{2}(\varrho_1+\varrho_2)=\phi_+^\star\phi_+
+\phi_-^\star\phi_-+\frac{1}{2}\Big(\mathcal{S}_{+1}^\star\mathcal{S}_{+1}+\mathcal{S}_{-1}^\star\mathcal{S}_{-1}
+\mathcal{S}_{+2}^\star\mathcal{S}_{+2}+\mathcal{S}_{-2}^\star\mathcal{S}_{-2}\Big)
\end{eqnarray}
we have
\begin{eqnarray}
\{\mathcal{C}^0(x),\varrho(x')\}=\{\mathcal{C}^\pm(x),\varrho(x')\}=0
\end{eqnarray}
and the following set of first class constraints
\begin{eqnarray}\label{2.17}
\mathcal{C}^0(x)\approx 0
\end{eqnarray}
\begin{eqnarray}\label{2.18}
\mathcal{C}^\pm(x)\approx 0
\end{eqnarray}
\begin{eqnarray}\label{2.19}
\varrho(x)\approx 1
\end{eqnarray}
ensures that on the constraint surface the decomposed Pauli spinor
(\ref{2.6}) reproduces the Poisson brackets (\ref{2.2}) of the
Pauli spinor.

Clearly, the condition (\ref{2.19}) is a generalization of
(\ref{1.6}) to the present case. It can be interpreted as a
statement that the decomposed liquid of spinons and chargons is
incompressible. There is also an obvious analogy between
(\ref{2.19}) and the Gaub law constraint in a $U(1)$ gauge theory.

Notice that for the bosonic fields the number operator
(\ref{2.16}) has eigenvalue one
\begin{eqnarray}
\{\varrho(x),\phi_{\pm}(x')\}=\phi_\pm(x)\delta(x-x')
\end{eqnarray}
but for the spinor components we get
\begin{eqnarray}
\{\varrho(x),\mathcal{S}_{\pm
\alpha}(x')\}=\frac{1}{2}\mathcal{S}_{\pm \alpha}(x)\delta(x-x').
\end{eqnarray}
This is since according to our normalization of $\varrho(x)$, the
two components of $\mathcal{S}_{\pm \alpha}$ together form one
spinor with particle number one. The condition (\ref{2.19}) states
that at each point we have one complete particle, either a chargon
or a spinon with its two components.

We also note that since (\ref{2.17})-(\ref{2.19}) is a first-class
algebra, once we introduce the subsidiary i.e. gauge fixing
conditions \cite{8}, each constraint leads to the elimination of a
pair of canonical variables. We then have four complex conditions
and as a consequence both sides of (\ref{2.6}) describe an equal
number of four independent field degrees of freedom: The
decomposition (\ref{2.6}) is complete and fully consistent with
the canonical structure of the original Pauli spinor.

When we compute the Poisson bracket of the constraint functionals
$\mathcal{C}^0, \mathcal{C}^\pm$ with the decomposed Pauli spinor,
we find that they act like the Pauli matrices,
\begin{eqnarray}
\{\mathcal{C}^0(x),\psi_1(x')\} &=&-\psi_1(x)\delta(x-x')
\end{eqnarray}
\begin{eqnarray}
\{\mathcal{C}^0(x),\psi_2(x')\}&=& \psi_2(x)\delta(x-x').
\end{eqnarray}
Let us now defined the Pauli spinor by $\psi=(\psi_1,\psi_2)$ and
the third component of Pauli matrix by $\sigma^3$, then
\begin{eqnarray}
\{\mathcal{C}^0(x),\psi(x')\}&=&-\left(\begin{array}{cc} 1&0\\0&-1
\end{array}\right)\left(\begin{array}{c}
\psi_1(x)\\
\psi_2(x)
\end{array}\right)\delta(x-x')\cr
&=&-\sigma^3\left(\begin{array}{c}
\psi_1(x)\\
\psi_2(x)
\end{array}\right)\delta(x-x')
\end{eqnarray}
\begin{eqnarray}
\{\mathcal{C}^+(x),\psi_1(x')\}&=&0
\end{eqnarray}
\begin{eqnarray}
\{\mathcal{C}^+(x),\psi_2(x')\}&=&\psi_1(x)\delta(x-x')
\end{eqnarray}
\begin{eqnarray}
\{\mathcal{C}^-(x),\psi_1(x')\} &=& \psi_2(x)\delta(x-x')
\end{eqnarray}
\begin{eqnarray}
\{\mathcal{C}^-(x),\psi_2(x')\}=0
\end{eqnarray}
 Note
that
$
\frac{1}{2}\Big(\sigma^1+i\sigma^2\Big)=\left(\begin{array}{cc}
0&1\\0&0
\end{array}\right)
$
and
$
\frac{1}{2}\Big(\sigma^1-i\sigma^2\Big)=\left(\begin{array}{cc}
0&0\\1&0
\end{array}\right).
$
Then
\begin{eqnarray}
\{\mathcal{C}^+(x),\psi(x')\}
&=&\frac{1}{2}\Big(\sigma^1-i\sigma^2\Big)\left(\begin{array}{c}
\psi_1(x)\\
\psi_2(x)
\end{array}\right)\delta(x-x')
\end{eqnarray}
\begin{eqnarray}
\{\mathcal{C}^-(x),\psi(x')\}
&=&\frac{1}{2}\Big(\sigma^1+i\sigma^2\Big)\left(\begin{array}{c}
\psi_1(x)\\
\psi_2(x)
\end{array}\right)\delta(x-x').
\end{eqnarray}
We conclude that
\begin{eqnarray}
\{\mathcal{C}^\pm(x),\psi(x')\}=\frac{1}{2}\Big(\sigma^1\mp
i\sigma^2\Big)\left(\begin{array}{c}
\psi_1(x)\\
\psi_2(x)
\end{array}\right)\delta(x-x').
\end{eqnarray}
For the density operator we get
\begin{eqnarray}
\{\varrho(x),\psi_1(x')\}&=&\frac{3}{2}\psi_1(x)\delta(x-x')
\end{eqnarray}
\begin{eqnarray}
\{\varrho(x),\psi_2(x')\}&=&\frac{3}{2}\psi_2(x)\delta(x-x')
\end{eqnarray}
Then
\begin{eqnarray}
\{\varrho(x),\psi(x')\}&=&\frac{3}{2}I\left(\begin{array}{c}
\psi_1(x)\\
\psi_2(x)
\end{array}\right)\delta(x-x'),
\end{eqnarray}
where the eigenvalue $3/2$ is consistent with our interpretation
of the condition (\ref{2.19}) with each component of the Pauli
spinor containing one full bosonic chargon and one of the two
components of a full spinon.

As in the case of the lattice fermion we again conclude that the
decomposition (\ref{2.6}) is not unique. For example, we may
introduce
\begin{eqnarray}\label{2.20}
\psi_\alpha= f(\phi,\phi^\star).\mathcal{S}_\alpha
\end{eqnarray}
and if we e.g. select $f$ to have the functional form
\begin{eqnarray}\label{2.21}
f(\phi,\phi^\star)=f(\frac{\phi}{\phi^\star});\,\,\,\,
f^\star(\phi,\phi^\star)=f(\phi^\star,\phi);\,\,\, |f|=1
\end{eqnarray}
which is a generalization of (\ref{1.17}), the decomposed spinor
obeys the Poisson brackets (\ref{2.2}). In the slave-fermion
approach, when we project the field $\psi(x)$ to a many-particle
subspace so that the chargon contribution is totally antisymmetric
under exchange of the particle coordinates and the spinon is
symmetric, this decomposition incorporates the (normalized)
Laughlin wavefunction \cite{11},\cite{7} of fractional quantum
Hall effect.
\subsection{Resolving the constraints}
We now proceed to explicitly resolve the constraints
(\ref{2.17}), (\ref{2.18}), (\ref{2.19}). For this we introduce a
tree component unit vector $\overrightarrow{s}=(s_1,s_2,s_3)$ i.e.
$(s_1^2+s_2^2+s_3^2=1)$. We denote
\begin{eqnarray}
s_\pm=s_1\pm is_2
\end{eqnarray}
and define two spinors $\chi_{\uparrow}$ and $\chi_{\downarrow}$
as follows,
\begin{eqnarray}
\chi_{\uparrow}=\left(\begin{array}{c} \chi_{\uparrow 1}
\\\chi_{\uparrow 2}
\end{array}\right)=\frac{1}{\sqrt{2(1-s_3)}}\left(\begin{array}{c}
1-s_3\\-s_+
\end{array}\right)
\end{eqnarray}
\begin{eqnarray}
\chi_{\downarrow}=\left(\begin{array}{c} \chi_{\downarrow 1}
\\\chi_{\downarrow 2}
\end{array}\right)=\frac{1}{\sqrt{2(1-s_3)}}\left(\begin{array}{c}
s_-\\1-s_3
\end{array}\right).
\end{eqnarray}
This defines an orthonormal basis of complex spinors,
\begin{eqnarray}
\chi_{\uparrow}^\dag.\chi_{\uparrow}&=&\frac{1}{2(1-s_3)}\Big(1-s_3,-s_-\Big)\left(\begin{array}{c}
1-s_3\\-s_+
\end{array}\right)=1
\end{eqnarray}
\begin{eqnarray}
\chi_{\downarrow}^\dag.\chi_{\downarrow}&=&\frac{1}{2(1-s_3)}\Big(s_+,1-s_3\Big)\left(\begin{array}{c}
s_-\\1-s_3
\end{array}\right)=1.
\end{eqnarray}
Then
\begin{eqnarray}\label{re1}
\chi_{\uparrow}^\dag.\chi_{\uparrow}=\chi_{\downarrow}^\dag.\chi_{\downarrow}=1.
\end{eqnarray}
In the same case
\begin{eqnarray}\label{re2}
\chi_{\uparrow}^\dag.\chi_{\downarrow}&=&\frac{1}{2(1-s_3)}\Big(1-s_3,-s_-\Big)\left(\begin{array}{c}
s_-\\1-s_3
\end{array}\right)=0.
\end{eqnarray}
We then can resume (\ref{re1}) and (\ref{re2}) as
\begin{eqnarray}
\chi_{\uparrow \alpha}^\dag \chi_{\uparrow
\beta}+\chi_{\downarrow\alpha}^\dag\chi_{\downarrow\beta}=\delta_{\alpha\beta}.
\end{eqnarray}
These two spinors are related to each other by a charge
conjugation
\begin{eqnarray}
-i\sigma^2\chi_{\uparrow}^\star&=&-i\left(\begin{array}{cc}
0&-i\\i&0
\end{array}\right)\frac{1}{\sqrt{2(1-s_3)}}\left(\begin{array}{c}
1-s_3\\-s_-
\end{array}\right)\cr
&=&\frac{1}{\sqrt{2(1-s_3)}}\left(\begin{array}{c} s_-\\1-s_3
\end{array}\right)=\chi_{\downarrow}.
\end{eqnarray}
This relation is equivalent to
\begin{eqnarray}
\left(\begin{array}{c} \chi_{\downarrow 1}
\\\chi_{\downarrow 2}
\end{array}\right)=\left(\begin{array}{cc} 0&-1\\1&0
\end{array}\right)\left(\begin{array}{c} \chi_{\uparrow 1}^\star
\\\chi_{\uparrow 2}^\star
\end{array}\right)\Leftrightarrow\Big(\chi_{\downarrow 1}=-\chi_{\uparrow
2}^\star,\,\,\,\,\,\chi_{\downarrow 2}=\chi_{\uparrow
1}^\star\Big).
\end{eqnarray}
Note that if we
introduce the spin projection operator
\begin{eqnarray}
\hat{s}=-\frac{1}{2}\overrightarrow{s}.\hat{\sigma}
\end{eqnarray}
where $\hat{\sigma}$ is the vector with Pauli matrices as components.
Then
\begin{eqnarray}
\hat{s}\chi_{\uparrow}&=&-\frac{1}{2}\Big(s_1\sigma_1+s_2\sigma_2+s_3\sigma_3\Big)\left(\begin{array}{c}
\chi_{\uparrow 1}
\\\chi_{\uparrow 2}
\end{array}\right)=\frac{1}{2}\chi_{\uparrow}.
\end{eqnarray}
In the same computation
\begin{eqnarray}
\hat{s}\chi_{\downarrow}&=&-\frac{1}{2}\frac{1}{\sqrt{2(1-s_3)}}\left(\begin{array}{cc}
s_3& s_-\\
s_+& -s_3
\end{array}\right)\left(\begin{array}{c}
s_-\\1-s_3
\end{array}\right)=-\frac{1}{2}\chi_{\downarrow}.
\end{eqnarray}
We conclude that the orthonormal spinors are the $\pm\frac{1}{2}$
eigenstates of $\hat{s}$
\begin{eqnarray}
\hat{s}\chi_{\uparrow}=\frac{1}{2}\chi_{\uparrow},\,\,\,\hat{s}\chi_{\downarrow}=-\frac{1}{2}\chi_{\downarrow}.
\end{eqnarray}
These relations can be inverted to represent $\overrightarrow{s}$
in terms of the spinors,
\begin{eqnarray}
\left(\begin{array}{cc} s_+\\s_3
\end{array}\right)=-\left(\begin{array}{cc} 2\chi_{\uparrow 1}^\star\chi_{\uparrow 2}\\
\chi_{\uparrow 1}^\star\chi_{\uparrow 1}-\chi_{\uparrow
2}^\star\chi_{\uparrow 2}
\end{array}\right)=+\left(\begin{array}{cc}
2\chi_{\downarrow 1}^\star\chi_{\downarrow 2}\\
\chi_{\downarrow 1}^\star\chi_{\downarrow 1}-\chi_{\downarrow
2}^\star\chi_{\downarrow 2}.
\end{array}\right)
\end{eqnarray}
This relation can be compressed in the form
\begin{eqnarray}\label{new}
\overrightarrow{s}&=&\left(\begin{array}{cc} s_+\\s_3
\end{array}\right)=-\chi_{\uparrow}^\dag\hat{\sigma}\chi_{\uparrow}=+\chi_{\downarrow}^\dag\hat{\sigma}\chi_{\downarrow}\cr
&=&-\left(\begin{array}{cc}
\chi_{\uparrow}^\dag\sigma^+\chi_{\uparrow}\\
\chi_{\uparrow}^\dag\sigma^3\chi_{\uparrow}
\end{array}\right)=\left(\begin{array}{cc}
\chi_{\downarrow}^\dag\sigma^+\chi_{\downarrow}\\
\chi_{\downarrow}^\dag\sigma^3\chi_{\downarrow}
\end{array}\right)
\end{eqnarray}
where $\sigma^+=\sigma_1+i\sigma_2=\left(\begin{array}{cc}
0&2\\0&0
\end{array}\right)$.

With these spinors we can resolve the constraints (\ref{2.17}),
(\ref{2.18}), (\ref{2.19}) as follows: We set
\begin{eqnarray}\label{zzz}
\mathcal{S}_+=\left(\begin{array}{cc} \mathcal{S}_{+1}
\\\mathcal{S}_{+2}
\end{array}\right)=:\rho_+.\chi_{\uparrow}
\end{eqnarray}
and
\begin{eqnarray}\label{xxx}
\mathcal{S}_-=\left(\begin{array}{cc} \mathcal{S}_{-1}
\\\mathcal{S}_{-2}
\end{array}\right)=:\rho_-.\chi_{\downarrow}
\end{eqnarray}
where
\begin{eqnarray}
|\rho_\pm|^2=\mathcal{S}_\pm^\dag\mathcal{S}_{\pm}.
\end{eqnarray}
When we substitute these in the constraints (\ref{2.17}) and
(\ref{2.18}) we get
\begin{eqnarray}
\mathcal{C}^0=\varrho_1-\varrho_2
=s_3\Big(|\rho_-|^2-|\rho_+|^2\Big)\approx 0
\end{eqnarray}
\begin{eqnarray}
\mathcal{C}^-&=&\mathcal{S}_{+1}^\star\mathcal{S}_{+2}+
\mathcal{S}_{-1}^\star\mathcal{S}_{-2}=\frac{1}{2}s_+\Big(|\rho_-|^2-|\rho_+|^2\Big)\approx
0
\end{eqnarray}
\begin{eqnarray}
\mathcal{C}^+&=&\mathcal{S}_{+2}^\star
\mathcal{S}_{+1}+\mathcal{S}_{-2}^\star
\mathcal{S}_{-1}=\frac{1}{2}s_-\Big(|\rho_-|^2-|\rho_+|^2\Big)\approx
0.
\end{eqnarray}
Then
\begin{eqnarray}
\mathcal{C}^\pm=\frac{1}{2}s_\mp\Big(|\rho_-|^2-|\rho_+|^2\Big)\approx
0.
\end{eqnarray}
Consequently, these constraints become resolved if we set
\begin{eqnarray}
\rho_\pm=r e^{i\omega_\pm},\quad r\in\mathbb{R}.
\end{eqnarray}
When we substitute the following decomposed spinor \cite{9}
\begin{eqnarray}\label{2.27}
\psi=\phi_+ r\chi_{\uparrow}(\overrightarrow{s})
e^{i\omega_+}+\phi_-
r\chi_{\downarrow}(\overrightarrow{s})e^{i\omega_-}
\end{eqnarray}
in the remaining constraint (\ref{2.19}), we get
\begin{eqnarray}
\varrho=\frac{1}{2}(\varrho_1+\varrho_2)&=&\phi_+^\star\phi_+
+\phi_-^\star\phi_-+\frac{1}{2}\Big(\mathcal{S}_{+1}^\star\mathcal{S}_{+1}
+\mathcal{S}_{-1}^\star\mathcal{S}_{-1}
+\mathcal{S}_{+2}^\star\mathcal{S}_{+2}+
\mathcal{S}_{-2}^\star\mathcal{S}_{-2}\Big)\cr
&=&\phi_+^\star\phi_+ +\phi_-^\star\phi_-+r^2\approx
1\cr
 &\Rightarrow&|\phi_+|^2+|\phi_-|^2\approx 1-r^2.
\end{eqnarray}
This defines a sphere $\mathbb{S}^2$, which we parametrize by
setting
\begin{eqnarray}
\left(\begin{array}{ccc} |\phi_+|\\
|\phi_-|\\r
\end{array}\right)=\left(\begin{array}{ccc} \cos\alpha\sin\beta\\
\sin\alpha\sin\beta\\\cos\beta
\end{array}\right)
\end{eqnarray}
and we define
\begin{eqnarray}
\phi_\pm=|\phi_\pm|e^{i\eta_\pm}.
\end{eqnarray}
For the Pauli spinor (\ref{2.27}), we then arrive at the following
explicit parametrization
\begin{eqnarray}
\psi&=&\left(\begin{array}{cc} \psi_1\\\psi_2
\end{array}\right)=\phi_+ r\chi_{\uparrow}(\overrightarrow{s})
e^{i\omega_+}+\phi_-
r\chi_{\downarrow}(\overrightarrow{s})e^{i\omega_-}\cr &=&
\cos\beta\Big(
\cos\alpha\sin\beta\chi_{\uparrow}(\overrightarrow{s})e^{i(\omega_+
+\eta_+)}
+\sin\alpha\sin\beta\chi_{\downarrow}(\overrightarrow{s})e^{i
(\omega_-+\eta_-)}\Big)\cr &=&\frac{1}{2}\sin
2\beta\Big(\cos\alpha\sin\chi_{\uparrow}(\overrightarrow{s})e^{i(\omega_+
+\eta_+)} +\sin\alpha\chi_{\downarrow}(\overrightarrow{s})e^{i
(\omega_-+\eta_-)}\Big).
\end{eqnarray}
Here the phase combinations $\omega_\pm+\eta_\pm$ parametrize the
internal $U(1)$ symmetries. If we choose the relative phases of
the spinons and chargons to cancel each other,
\begin{eqnarray}
\omega_\pm+\eta_\pm=
\end{eqnarray}
we obtain
\begin{eqnarray}\label{2.28}
\psi=:\phi_{\uparrow}\chi_{\uparrow}+\phi_{\downarrow}\chi_{\downarrow}
\end{eqnarray}
with only four field degrees of freedom in (\ref{2.28}) that is consistent
with the constraint structure. However, here we prefer to leave
the relative phases between the spinons and chargons as
unspecified. We wish to treat the spinon fields and the chargon
fields as independent dynamical degrees of freedom.
\label{subsect3}

We now come to the problem with such a decomposition, and the
question is how to resolve it: For the Pauli spinor the condition
(\ref{2.19}) yields
\begin{eqnarray}
|\psi|^2&=&\Big(\phi_+ r\chi_{\uparrow}(\overrightarrow{s})
e^{i\omega_+}+\phi_-
r\chi_{\downarrow}(\overrightarrow{s})e^{i\omega_-}\Big)\Big(
\phi_+ r\chi_{\uparrow}(\overrightarrow{s}) e^{i\omega_+}+\phi_-
r\chi_{\downarrow}(\overrightarrow{s})e^{i\omega_-}\Big)^\dag\cr
&=& r^2(|\phi_+|^2+|\phi_-|^2)=r^2(1-r^2).
\end{eqnarray}
Note that this vanishes when either the density of spinons ($r=0$)
or the density of chargons ($r=1$) vanishes, and that the maximum
value is obtained at $r=1/\sqrt{2}$ which gives for the maximum
value of the absolute value of spinor
\begin{eqnarray}
0\leq |\psi|\leq \frac{1}{2}.
\end{eqnarray}
This relation shows that the construction is not complete: For a
Schrodinger equation we demand that the integral of $|\psi|^2$ is equal to $=1,$
but at a given point $x$ the absolute value of the wavefunction
can be any real number.
\section{New class of  decomposition:  resolution  of the constraint equations}
In this section, I show that not all decompositions are
possible. 

$\bullet$ Let
us define the spin-charge decomposition as follows:
\begin{eqnarray}
\psi_\alpha(x)=\phi_+\mathcal{S}_{+\alpha}+\epsilon_{\alpha\beta}\phi_-\mathcal{S}_{-\beta}
\end{eqnarray}
with the Poisson brackets of the fields $\phi_\pm$ and
$\mathcal{S}_{l\alpha}$ as
\begin{eqnarray}
\{\phi_\pm^\star(x),\phi_\pm(x')\}=\pm\delta(x-x'),\,\,\,\,
\{\mathcal{S}_{\pm\alpha}^\star(x),\mathcal{S}_{\pm\beta}(x')\}=\pm\delta_{\alpha\beta}(x-x')
\end{eqnarray}
and
\begin{eqnarray}
\{\phi_\pm^\star(x),\phi_\mp(x')\}=0,\,\,\,\,
\{\mathcal{S}_{\pm\alpha}^\star(x),\mathcal{S}_{\mp\beta}(x')\}=0.
\end{eqnarray}
There result the Poisson brackets
\begin{eqnarray}
\{\psi_1^\star(x),\psi_1(x')\}&=&\Big(\phi_+^\star\phi_+-\phi_-^\star\phi_-
+\mathcal{S}_{+1}^\star\mathcal{S}_{+1}-\mathcal{S}_{-2}^\star\mathcal{S}_{-2}\Big)(x)\delta(x-x')\cr
&=:&\varrho_1(x)\delta(x-x')
\end{eqnarray}
\begin{eqnarray}
\{\psi_2^\star(x),\psi_2(x')\}&=&\Big(\phi_+^\star\phi_+-\phi_-^\star\phi_-
+\mathcal{S}_{+2}^\star\mathcal{S}_{+2}-\mathcal{S}_{-1}^\star\mathcal{S}_{-1}\Big)(x)\delta(x-x')\cr
&=:&\varrho_2(x)\delta(x-x')
\end{eqnarray}
\begin{eqnarray}
\{\psi_1^\star(x),\psi_2(x')\}=\Big(\mathcal{S}_{+1}^\star\mathcal{S}_{+2}+\mathcal{S}_{-2}^\star\mathcal{S}_{-1}\Big)
(x)\delta(x-x')=:\mathcal{C}^-(x)\delta(x-x')
\end{eqnarray}
\begin{eqnarray}
\{\psi_2^\star(x),\psi_1(x')\}=\Big(\mathcal{S}_{+2}^\star\mathcal{S}_{+1}+\mathcal{S}_{-1}^\star\mathcal{S}_{-2}\Big)
(x)\delta(x-x')=:\mathcal{C}^+(x)\delta(x-x')
\end{eqnarray}
Then
\begin{eqnarray}
\{\mathcal{C}^-(x),\mathcal{C}^+(x')\}=\Big(\varrho_2-\varrho_1\Big)(x)\delta(x-x')=:\mathcal{C}^0(x)\delta(x-x').
\end{eqnarray}
and
\begin{eqnarray}
\{\mathcal{C}^+(x),\mathcal{C}^0(x')\}=2\mathcal{C}^+(x)\delta(x-x'),\,\,\,\,\{\mathcal{C}^-(x),\mathcal{C}^0(x')\}=-2\mathcal{C}^-(x)\delta(x-x').
\end{eqnarray}
showing that $(\mathcal{C}^\pm, \mathcal{C}^0)$ determine an
$SU(2)$ algebra. Besides,
\begin{eqnarray}
\{\mathcal{C}^+(x),\varrho_1(x')\}=-\{\mathcal{C}^+(x),\varrho_2(x')\}=-\mathcal{C}^+(x)\delta(x-x')
\end{eqnarray}
\begin{eqnarray}
\{\mathcal{C}^-(x),\varrho_1(x')\}=-\{\mathcal{C}^-(x),\varrho_2(x')\}=\mathcal{C}^-(x)\delta(x-x').
\end{eqnarray}
If we define the total density operator as
\begin{eqnarray}
\varrho=\frac{1}{2}(\varrho_1+\varrho_2)=\phi_+^\star\phi_+-\phi_-^\star\phi_-
+\frac{1}{2}\Big(\mathcal{S}_{+1}^\star\mathcal{S}_{+1}-\mathcal{S}_{-2}^\star\mathcal{S}_{-2}+
\mathcal{S}_{+2}^\star\mathcal{S}_{+2}-\mathcal{S}_{-1}^\star\mathcal{S}_{-1}\Big)\
\end{eqnarray}
we have
\begin{eqnarray}
\{\mathcal{C}^0(x),\varrho(x')\}=\{\mathcal{C}^\pm(x),\varrho(x')\}=0
\end{eqnarray}
and the following set of first class constraints
\begin{eqnarray}\label{dineb}
\mathcal{C}^0(x)\approx 0,\,\,\, \mathcal{C}^\pm(x)\approx
0,\,\,\, \varrho(x)\approx 1
\end{eqnarray}
ensuring that on the constraint surface the decomposed Pauli spinor
reproduces the Poisson brackets (\ref{2.2}).

We have to solve the constraint equation (\ref{dineb}). By expansion we get
\begin{eqnarray}
\mathcal{C}^0=\mathcal{S}_{+2}^\star\mathcal{S}_{+2}-\mathcal{S}_{-1}^\star\mathcal{S}_{-1}
+\mathcal{S}_{+1}^\star\mathcal{S}_{+1}-\mathcal{S}_{-2}^\star\mathcal{S}_{-2}\approx
0.
\end{eqnarray}
Recall that
\begin{eqnarray}
\mathcal{S}_+^\dag
\sigma^3\mathcal{S}_+=\mathcal{S}_{+1}^\star\mathcal{S}_{+1}-\mathcal{S}_{+2}^\star\mathcal{S}_{+2},\,\,\,\,
\mathcal{S}_-^\dag
\sigma^3\mathcal{S}_-=\mathcal{S}_{-1}^\star\mathcal{S}_{-1}-\mathcal{S}_{-2}^\star\mathcal{S}_{-2}.
\end{eqnarray}
Then
\begin{eqnarray}\label{con1}
\mathcal{C}^0&=&-\mathcal{S}_+^\dag \sigma^3\mathcal{S}_+
-\mathcal{S}_-^\dag \sigma^3\mathcal{S}_- = s_3(|\rho_+|^2-|\rho_-|^2)\approx 0
\end{eqnarray}
\begin{eqnarray}
\mathcal{C}^-=\mathcal{S}_{+1}^\star\mathcal{S}_{+2}+\mathcal{S}_{-2}^\star\mathcal{S}_{-1},\,\,\,
\mathcal{C}^+=\mathcal{S}_{+2}^\star\mathcal{S}_{+1}+\mathcal{S}_{-1}^\star\mathcal{S}_{-2}.
\end{eqnarray}
Besides,
\begin{eqnarray}
&&\mathcal{S}_+^\dag
\sigma^+\mathcal{S}_+=2\mathcal{S}_{+1}^\star\mathcal{S}_{+2},\,\,\,\,\mathcal{S}_-^\dag
\sigma^+\mathcal{S}_-=2\mathcal{S}_{-1}^\star\mathcal{S}_{-2},\\
&& \mathcal{S}_+^\dag
\sigma^-\mathcal{S}_+=2\mathcal{S}_{+2}^\star\mathcal{S}_{+1},\,\,\,\,\mathcal{S}_-^\dag
\sigma^-\mathcal{S}_-=2\mathcal{S}_{-2}^\star\mathcal{S}_{-1}
\end{eqnarray}
yielding
\begin{eqnarray}
\chi_\downarrow^\dag
\sigma^-\chi_\downarrow=s_-=-\chi_\uparrow^\dag
\sigma^-\chi_\uparrow.
\end{eqnarray}
Then we get the following results
\begin{eqnarray}\label{con2}
\mathcal{C}^-=\frac{1}{2}\Big(-|\rho_+|^2 s_++|\rho_-|^2
s_-\Big)\approx 0,\\
\mathcal{C}^+=\frac{1}{2}\Big(-|\rho_+|^2 s_-+|\rho_-|^2
s_+\Big)\approx 0,
\end{eqnarray}
and
\begin{eqnarray}\label{con3}
\varrho&=&
|\phi_+|^2-|\phi_-|^2+\frac{1}{2}\Big(\mathcal{S}_+^\dag\mathcal{S}_+-\mathcal{S}_-^\dag\mathcal{S}_-\Big)\cr
&=&|\phi_+|^2-|\phi_-|^2+\frac{1}{2}\Big(|\rho_+|^2-|\rho_-|^2\Big)\approx
1.
\end{eqnarray}
The equations (\ref{con1}) and (\ref{con3}) show that
\begin{eqnarray}
|\phi_+|^2-|\phi_-|^2\approx 1
\end{eqnarray}
which can be solved in the constraint surface to give
\begin{eqnarray}
\phi_+=e^{i\nu_+}\cosh\alpha ,\,\,\,\,\,
\phi_-=e^{i\nu_-}\sinh\alpha
\end{eqnarray}
where $\alpha$ and $\nu_\pm$ are  real parameters.

If we choose $\rho_\pm=re^{i\omega_\pm}$, (\ref{con1}) can be
easily solved and the relations  (\ref{con2}) give
\begin{eqnarray}
\mathcal{C}^\pm=\pm\frac{1}{2}r^2(s_+-s_-)\approx 0 \Rightarrow
s_+\approx s_-\Rightarrow s_2\approx 0
\end{eqnarray}
The absolute value of field $\psi$ is then given by
\begin{eqnarray}
|\psi|=r,\,\mbox{ and }\, 0\leq |\psi| <\infty.
\end{eqnarray}

$\bullet$ We
adapt here the following spin charge decomposition
\begin{eqnarray}
\psi_\alpha=\frac{1}{2}\Big(\phi_+\mathcal{S}_{+\alpha}+\phi_+^\star\mathcal{S}_{+\alpha}+\epsilon_{\alpha\beta}\phi_-\mathcal{S}_{-\beta}
+\epsilon_{\alpha\beta}\phi_-^\star\mathcal{S}_{-\beta} \Big).
\end{eqnarray}
One can  show that
\begin{eqnarray}
\{\psi_1^\star(x),\psi_1(x')\}=\Big(\mathcal{R}^2_e(\phi_+)-\mathcal{R}^2_e(\phi_-)\Big)(x)\delta(x-x')=\{\psi_2^\star(x),\psi_2(x')\}
\end{eqnarray}
and
\begin{eqnarray}\label{5.3}
\varrho_1=\varrho_2=\mathcal{R}^2_e(\phi_+)-\mathcal{R}^2_e(\phi_-)\approx
1.
\end{eqnarray}
Beside
\begin{eqnarray}
\{\psi_1^\star(x),\psi_2(x')\}=\{\psi_2^\star(x),\psi_1(x')\}=0.
\end{eqnarray}
We now assumed that
$
\mathcal{R}^2_e\phi_-=\lambda.
$
Then (\ref{5.3}) give
$
\mathcal{R}^2_e\phi_+\approx 1+\lambda.
$

the function  $\rho_-$ and $\rho_+$  defined in (\ref{zzz}) and (\ref{xxx}) are  expressed as:
\begin{eqnarray}
\rho_+=re^{i\theta_+},\,\,\,\,\rho_-=(1-r)e^{i\theta_-}.
\end{eqnarray}
finally we conclude that
\begin{eqnarray}
|\psi|^2&=& r^2 \mathcal{R}^2_e(\phi_+) +(1-r)^2
\mathcal{R}^2_e(\phi_-)= r^2 +\Big(2r^2-2r+1\Big)\lambda
\end{eqnarray}
and
\begin{eqnarray}
|\psi|=\sqrt{r^2 +\Big(2r^2-2r+1\Big)\lambda}.
\end{eqnarray}
We arrived to the bound
\begin{eqnarray}
\sqrt{\frac{\lambda(\lambda+1)}{2\lambda+1}}\leq |\psi|< \infty.
\end{eqnarray}

Let us remark that in general case 
we can  choose arbitrary function
$f(r)$ and $g(r)$ such that
\begin{eqnarray}
\rho_+=f(r)e^{i\theta_+},\,\,\,\,\rho_-=g(r)e^{i\theta_-}
\end{eqnarray}
 and then
\begin{eqnarray}
|\psi|=\sqrt{f(r)^2+\Big(f(r)^2+g(r)^2\Big)\lambda}.
\end{eqnarray}
We defined  $r_0^i, i=1,2\cdots$ as the  solution of
$
(1+\lambda)f'(r)f(r)+\lambda g'(r)g(r)=0
$
such that
\begin{eqnarray}
\sqrt{f(r_0^i)^2+\Big(f(r_0^i)^2+g(r_0^i)^2\Big)\lambda}\leq
|\psi|<\infty.
\end{eqnarray}
For example,  setting
\begin{eqnarray}
f(r)=e^{kr^n},\,\,\,\,
g(r)=e^{-\frac{1+\lambda}{\lambda}kr^n};\,\,\,n\in\mathbb{N}
\end{eqnarray}
and 
\begin{eqnarray}
\forall r>0, \sqrt{f(r)^2+\Big(f(r)^2+g(r)^2\Big)\lambda}\leq
|\psi|<\infty .
\end{eqnarray}

\section*{Acknowledgments}
I indebted to Antti Niemi for having proposed
us the problem treated here. Discussion with Mahouton Norbert Hounkonnou
 is gratefully acknowledged. 
This work is partially supported by the ICTP through the OEA-ICMPA-Prj-
15. The ICMPA is in partnership with the Daniel Iagolnitzer Foundation (DIF),
France.


\begin{thebibliography}{99}

\bibitem{spint}
S.A. Wolf {\it et al}, Science {\bf 294}, 1488 (2001);
P. Sharma,
{\it ibid.}
{\bf 307}, 531 (2005)

\bibitem{fad1} L.D.~Faddeev, A.J.~Niemi,
  Phys.\ Lett.\ B {\bf 525}, 195 (2002).

\bibitem{9}
M.N. Chernodub and Antti J. Niemi, JETPLett.  {\bf 85} (2007) 
(quant-ph/0604162).

\bibitem{1}
P.A. Lee, N. Nagaosa and X.-G. Wen, {\it Rev. Mod. Phys. } {\bf
78} 17 (2006) 17;

E. Dagotta, {\it Rev. Mod. Phys. } {\bf 66} (1994) 000763.

\bibitem{2}
N. Nagaosa, {\it Quantum Field Theory in Strongly Correlated
Electronic Systems 9 Springer Verlag, Berlin} (1994).




\bibitem{3}
T.H. Hanson, {\it Phys. Repts.} {\bf 398} (2004) 327.

\bibitem{4}
 B.J Kim, H. Koh, E. Rotenberg, S.-J. Oh, H. Eisaki, N. Motoyama,
 S. Uchida, T. Tohyama, S. Maekawa, Z.-X. Shen and C. Kim, {\it Nature
 Physics} {\bf 2} (2006) 397.

 \bibitem{5}

J. Hubbard, {\it Proc. Roy. Soc.} {\bf 276} (1963)  238

E.H. Lieb, {\it XI Int. Cong. MP, Int. Press,} p.392 (1995)
(cond-mat/9311033).

\bibitem{takhta} L.D. Faddeev, L.A. Takhtajan, Phys. Lett. {\bf A85},  375 (1981).

\bibitem{andersson}
G. Baskaran, P.W. Anderson,
  Phys.\ Rev.\ {\bf B37}, 580 (1988).

\bibitem{7}
A. Lerda, {\it Anyons: Quantum mechanics of particles with
fractional spin} (Lecture notes in Physics), (Springer Verlag,
Berlin, 1993).

\bibitem{8}
P.A.M. Dirac, {\it Lectures on Quantum Mechanics} (Dover Books on
Physics, (2001)).


\bibitem{11}
R.B. Laughling, {\it Phys. Rev. Lett.} {\bf 50} (1983) 1395.


\bibitem{fqhe} F. Wilczek,
{\it Fractional Statistics and Anyon Superconductivity}
(World Scientific, Singapore, 1990).


\end{thebibliography}
\end{document}